\documentclass{svjour3}                     
\smartqed  
\usepackage{graphicx}
\usepackage{mathptmx}      
\usepackage{latexsym}
\newcommand{\msun}{{\rm M_{\odot}}}
\newcommand{\et}{{et al}}
\newcommand{\chandra}{{Chandra}}
\newcommand{\xmm}{{XMM-Newton}}
\newcommand{\suzaku}{{Suzaku}}

\begin{document}

\title{Searching for Black Holes in Space}
\subtitle{- the key role of x-ray observations}

\author{Ken Pounds}

\institute{Department of Physics and Astronomy, University of Leicester, UK\\
              Tel.: +44-116-2523509\\
              Fax: +44-116-2523311\\
              \email{kap@le.ac.uk}}

\date{Received: date / Accepted: date}

\maketitle

\begin{abstract}
Although General Relativity had provided the physical basis of black holes, evidence for their existence had to await the Space Era when X-ray
observations first directed the attention of astronomers to the unusual binary stars Cygnus X-1 and A0620-00. Subsequently, a number of faint Ariel 5 and
Uhuru X-ray sources, mainly at high Galactic latitude, were found to lie close to bright Seyfert galaxies, suggesting the nuclear activity in AGN might
also be driven by accretion in the strong gravity of a black hole. Detection of rapid X-ray variability with EXOSAT later confirmed that the accreting
object in AGN was almost certainly a supermassive black hole.    
\end{abstract}

\section{Introduction}
\label{intro}

The first recorded suggestion that there may be stars too massive for light to escape from their surface appears to have been made in  1783 by John
Michell (McCormack 1968, Hockey 2007), a rector working at a church near Leeds in Yorkshire. Michell, who had studied geology at Cambridge, but chose a better-remunerated position in
the church, also proposed that binary stars must have a common origin (rather than by chance encounter) and hence seems a worthy first  reference in this
historical review. However, a physical basis for the implied 'light bending' had to await developments in general relativity by Einstein and
Schwarzschild more than a century later. 

A key to the subsequent discovery of black holes, a term proposed by John Wheeler in 1965, is the extremely small (Schwarzschild)  radius of the
event horizon, with a correspondingly large potential energy release from accreting matter prior to falling into the hole. A further consequence of
the small emission region is that the radiation temperature will be high, indicating that much of the luminosity will lie in the far UV or X-ray
wavebends. 

Such simple considerations make it clear, with hindsight, why the first evidence that black holes actually exist had to await the Space Era, when the
first opportunity arose to send X-ray instruments above the Earth's atmosphere. In the event, although a number of remarkably bright non-solar X-ray
sources were discovered from brief sounding rockets flights during the 1960s, convincing evidence for the existence of stellar black holes had
to await the launch of the first dedicated satellites, in particular Uhuru and Ariel 5, launched in 1970 and 1974. 

Theoretical developments through the 1960s were slowed by the uncertain nature of most cosmic X-ray sources, though the binary star identification of
Sco X-1 was a strong stimulus to accretion disc theory (Shakura and Sunyaev 1973). Accretion onto a massive object in the  nucleus of an active galaxy (AGN) was 
considered by Salpeter (1964) and by Zel'dovich and Novikov (1964), and developed with the benefit of more efficient disc accretion by Lynden-Bell
(1969). However, it was only after the identification of Cygnus X-1 as a strong black hole candidate that a comparable stellar model was developed by Pringle
and Rees (1972) and by Shakura and Sunyaev (1973). 

It is hoped that this - necessarily - personal account will provide a useful review of one of the most significant developments in astrophysics during the Space Era.

\section{The origins of X-ray Astronomy}

Prior to the historic discovery (Giacconi et al 1962) of a remarkably bright X-ray source in the constellation of Scorpius, in June 1962, most
astronomers considered that observations in the ultraviolet and gamma ray bands offered the best promise for exploiting the exciting potential of space
research. X-ray observations were expected to focus on the study of active stars, with fluxes scaled from that of the solar corona, the only
known X-ray source at that time. Predictions ranged up to a thousand times the Sun's X-ray luminosity, but seemed beyond the reach of detection with
then-current technology. 

In a reflection of the contemporary thinking the recently formed US National Aeronautics and Space Agency (NASA) was planning a
series of Orbiting Astronomical Observatories, with the first missions devoted to UV astronomy, although a proposal in 1961 from the
UCL and Leicester groups to make simultaneous X-ray observations of the primary UV targets was accepted, and eventually flown on OAO-3
(Copernicus) eleven years later. In the USA, Riccardo Giacconi at American Science and Engineering (ASE) and Bruno Rossi of MIT had, still earlier, published the design of a more ambitious imaging
X-ray telescope, with nested mirrors for increased throughput (Giacconi and Rossi 1960), while Rossi made what was to prove a visionary statement in
declaring that 'Nature so often leaves the most daring imagination of man far behind'. 

The ASE/MIT Aerobee 150 sounding rocket flight from the White Sands Missile Range in June 1962, finding in Sco X-1 a cosmic X-ray
source a million time more luminous than the Sun (and brighter than the non-flaring corona at a few keV), began a transformation that laid 
the foundations for a revolution in High Energy Astrophysics.

Further sounding rocket observations quickly followed, with the US Naval Laboratory group - responsible for still earlier, but unrewarded night-time flights
(Friedman 1959) - confirming Sco X-1 and finding a further source in Taurus (Bowyer \et\ 1964a). The AS\&E group (Gursky \et\ 1963), and a team at
Lockheed (Fisher \et\ 1966) continued with launches from the White Sands missile range, while the British Skylark
rocket was used to explore the southern sky from Woomera in South Australia from 1967 (Harries \et\ 1967, Cooke \et\ 1967), finding several sources in Centaurus and the Galactic
centre region, exhibiting both thermal and non-thermal X-ray spectra (Cooke \et\ 1971). With a hint of what lay ahead, repeated Skylark observations found
the X-ray flux from Cen X-3 (first seen by the Lawrence Livermore Group; Chodil \et\ 1967) varied by an order of magnitude, while Cen X-2
briefly outshone Sco X-1, before apparently disappearing. 

Most sources remained of unknown nature, a notable exception being Tau X-1 which the NRL  group had identified with extended X-ray emission from the Crab Nebula supernova remnant in a classic use
of the Moon as an occulting  disc (Bowyer \et\ 1964b). Development of the modulation collimator (Oda \et\ 1965), with an important
refinement by Gursky, then yielded an arc minute position for
Sco X-1 (Gursky \et\ 1966), enabling its optical identification with a 13th magnitude blue star (Sandage et al 1966). By the end of the decade some
20-30 cosmic X-ray sources had been reliably detected, the uncertainty being partly due to many cosmic X-ray sources being highly variable or transient.
However, the majority remained unidentified, a major problem being their poorly determined positions on the sky.

\begin{figure}
\begin{center}
\hbox{
 \hspace{1.75 cm}
\includegraphics[width=8 cm]{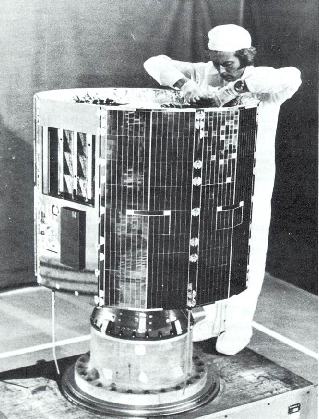}}
\end{center}
\caption{Ariel 5 spacecraft. The Sky Survey Instrument (SSI) detector array is seen on the upper left}
\label{fig:1}      
\end{figure}

The first small orbiting satellites dedicated to observations of cosmic X-ray sources began a transformation which quickly led to X-ray astronomy
becoming a major branch of astrophysics. Uhuru led the way in December 1970, being launched into a low equatorial orbit and carrying a large
proportional counter array to undertake a deep all-sky survey. Within a few months the extended observations made possible from orbit had shown that
many X-ray sources were variable. This led to the important finding that Cen X-3 and Her X-1 - and by implication probably many of the most luminous 
galactic sources - 
were binary star systems. Moreover, the rapid and periodic X-ray variations of Cen X-3 and Her X-1 showed the X-ray component to be most likely a neutron star.

A second remarkable Uhuru discovery - of extended X-ray emission from galaxy clusters quickly followed - while the number of known X-ray sources increased by an
order of magnitude. The 3U
catalogue (Giacconi \et\ 1974), listing 161 sources, was a key milestone in the development of X-ray astronomy, and the major scientific impact
of Uhuru is well recorded in Giacconi  and Gursky (1974). The identification of Cygnus X-1 and its recognition as the first strong candidate to contain
a black hole is described in more detail in Section 3.

Other Uhuru-class satellites followed, with Ariel 5 (UK), SAS-3 (USA) and Hakucho (Japan) dedicated to X-ray observations and OSO-7 (USA) and ANS
(Netherlands) being solar and UV astronomy missions carrying secondary X-ray instrumentation. For astronomers in the UK, Ariel 5 brought an ideal
opportunity to play a part in the rapid advances taking place. Like Uhuru, Ariel 5 (figure 1) was launched on a Scout rocket into a circular near-Earth
orbit from a disused oil platform off the coast of Kenya. It carried 6 experiments, the most important in the context of this paper being a Sky Survey
Instrument (SSI), similar to that on Uhuru, and an All Sky Monitor (from Goddard Space Fight Center). The Ariel 5 orbit was a good choice, not only in
minimising background due to cosmic rays and trapped radiation, but in allowing regular data dumps from the small on-board data recorder. A direct
ground and satellite link to the UK provided 6 orbits of `Quick Look' data from the SSI within an hour of  ground station
contact. The remaining `bulk' data were received within 24 hours, an immediacy  that contributed substantially to the excitement of the mission
operations, while also ensuring a rapid response to new discoveries. 

One such discovery was particularly well-timed, with the SSI detecting a previously unseen source in the constellation Monoceros, just 2 days before
the  start of the first European Astronomy Society meeting, held in Leicester, where new X-ray results from the Ariel 5 and SAS-3  missions dominated
the programme. The new X-ray source was to become a further strong black hole candidate, as reported in Section 4. In all, some 27 soft X-ray transients were
discovered as the Ariel 5 mission continued until 1980, though none were as powerful as A0620-00, the majority being neutron star binaries. A second 
enduring result from the Ariel 5 SSI was in establishing powerful X-ray emission as a characteristic property of Seyfert galaxies. The challenge of correctly identifying many previously unidentified sources,
individually located to a few tenths of a sq. deg, was possible only because Seyfert nuclei are also unusually bright in the optical band. While the
initial classification was made on a statistical basis, it held up well as new data emerged to establish Seyfert galaxies as
a major class of extragalactic X-ray source (Section 5). 

\section{Cygnus X-1}

Cygnus X-1 was first detected (figure 2) in an NRL Aerobee launch from White Sands in New Mexico in 1964 (Bowyer \et\ 1965). Subsequent observations, including a
balloon-borne telescope (Overbeck and Tananbaum 1968), showed an unusually hard power law spectrum, not unlike the Crab Nebula, but the lack of an obvious radio or
optical counterpart (within the large position uncertainty typical of those early detections) limited detailed investigation.  

\begin{figure}
\begin{center}
\hbox{
 \hspace{0.75 cm}
\includegraphics[width=10 cm]{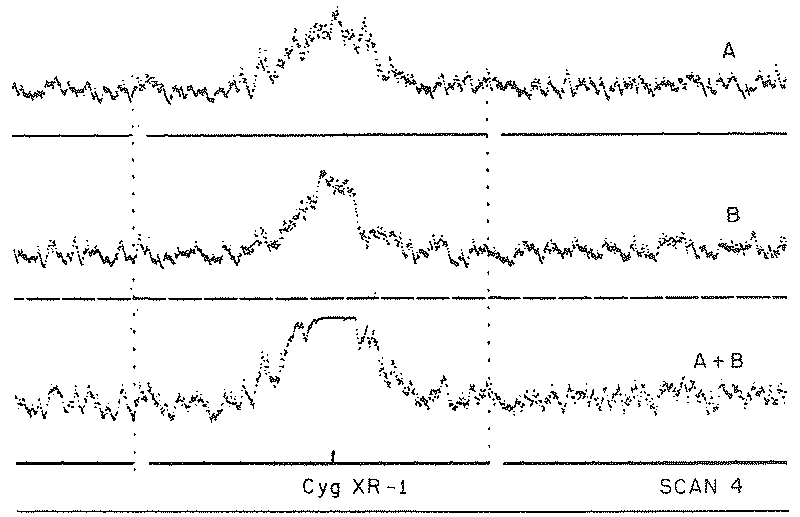}}
\end{center}
\caption{Telemetry traces from the NRL rocket-borne Geiger counters which first detected Cygnus X-1 in 1964 (Bowyer \et\ 1965)}     
\end{figure}

The unusual nature of Cygnus X-1 became clearer from extended Uhuru observations (figure 3) which showed large amplitude fluctuations in the X-ray  intensity on timescales down to
100 ms (Schreier \et\ 1971), implying a correspondingly small emission region.  
An improved X-ray source location from Uhuru and an MIT rocket flight (Rappaport \et\ 1971) led to the discovery of a weak transient radio source by 
Braes and Miley (1971) from Leiden Observatory, and independently by Hjellming and Wade (1971) at the NRAO, which subsequent Uhuru analysis showed to
coincide with a change in the X-ray appearance.  The greatly improved source position (figure 4) finally allowed the optical identification of
`Cyg X-1 - a Spectroscopic Binary with a Heavy companion'  by Webster and Murdin (1972), at the Royal Greenwich Observatory, and Bolton (1972),
at the  David Dunlap Observatory in Toronto.

\begin{figure}
\hbox{
 \hspace{0.75 cm}
\includegraphics[width=10 cm]{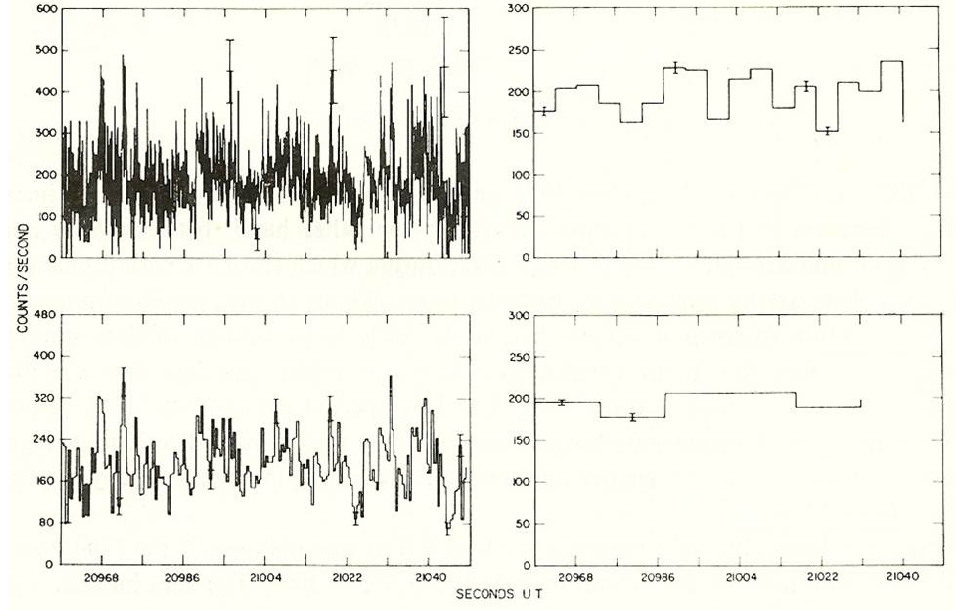}}
\caption{Uhuru observation of rapid non-periodic variability in Cygnus X-1 binned at 0.096s, 0.48s, 4.8s and 14.4s. Typical 1$\sigma$ error
bars are shown (Schreier \et\ 1971)}    
\end{figure}

\begin{figure}
\hbox{
\hspace{0.75 cm}  
\includegraphics[width=10cm]{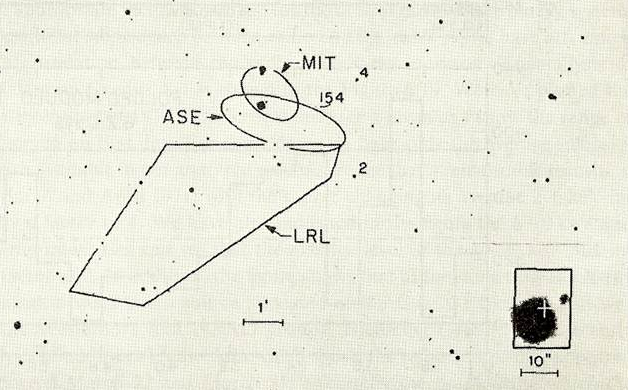}}
\caption{X-ray observations of Cygnus X-1. HDE 226868 is the bright star in the overlap of the ASE and MIT error boxes.
The insert shows the more precise coincidence of HDE 226868 with a transient radio source}
\end{figure}

The `heavy' companion was HDE 226868, a supergiant star at some 2 kpc distance. 
Notwithstanding a likely mass of $\sim$30 $\msun$, optical spectroscopy showed Doppler shifts of amplitude $\sim$65 km s$^{-1}$ which, together with a
binary period of 5.6 days, indicated the unseen (X-ray) companion to have a mass of $\sim$15 $\msun$, much larger than the maximum value expected for a
neutron star. Cygnus X-1 became widely - though not universally - acclaimed as the first stellar black hole (eg Shipman 1975), and understanding the nature of Cygnus X-1 was a primary
motivation of theoretical work on accretion discs during that period (see Pringle 1977 for a contemporary review).

VLBA observations in 2009-10 (Reid \et\ 2011) finally resolved the long-standing uncertainty in the distance of Cygnus X-1, obtaining a  parallax distance of
1.86$\pm$0.12 kpc.   Re-analysis of extensive X-ray and optical data using the new value has allowed a refinement of the black hole mass of 14.8$\pm$1.0
$\msun$ (Orosz \et\ 2011),  and shown that Cygnus X-1 contains a near-extreme Kerr black hole with a spin parameter $a_*>0.95$, corresponding to a spin
rate of $\sim$800 s$^{-1}$ (Gou \et\ 2011).

\section{A0620-00}

A0620-00 (=V616 Mon) was much the brightest of several X-ray transients discovered by the Ariel 5 satellite during an extended scan of the Galactic plane in autumn
1975. Within 3 days of its first sighting it was  brighter
than the Crab Nebula, while 2 days later it outshone Scorpius X-1, becoming - for a few weeks - the brightest cosmic X-ray source ever seen (Elvis \et\  1975),
a record to be held for 30 years. Well before peaking at a flux level $\sim$3 times that of Sco X-1, the new source was being monitored 
by Ariel 5, SAS-3 and other space- and ground-based telescopes around the world. The X-ray emission then faded over several months, being
continually monitored by the Goddard All Sky Monitor on Ariel 5 (Kaluzienski \et\ 1975).

\begin{figure*}
\begin{center}
\hbox{
 \hspace{0.25 cm}
  \includegraphics[width=7 cm]{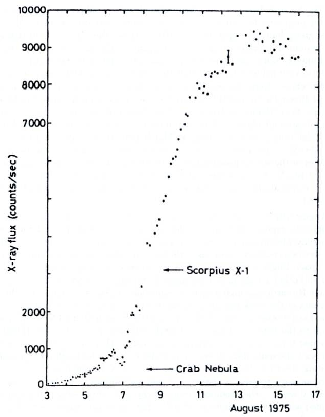}
 \hspace{0.3 cm}
  \includegraphics[width = 3.6 cm]{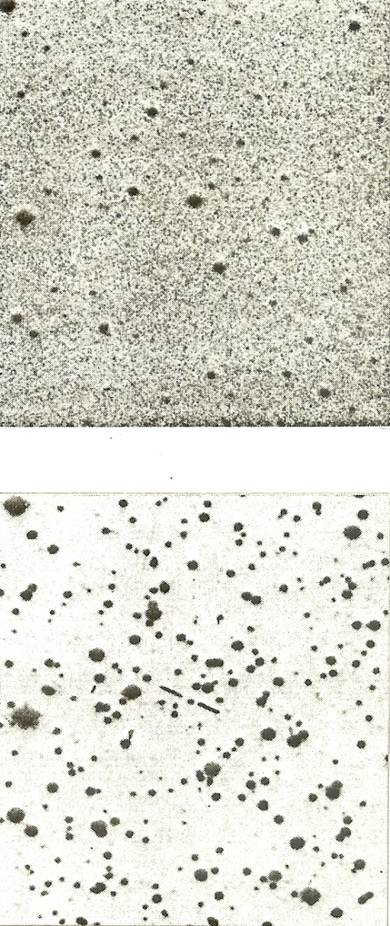}}
  \end{center}
\caption{(left)Soft X-ray transient source A0620-00 (Nova Mon) detected in an Ariel 5 Galactic Plane survey. (right) Comparison of a short UK Schmidt telescope 
exposure taken during outburst (top) with the corresponding Palomar Sky Survey red plate (lower), from which Boley and Wolfson (1975) identified A0620-00 with the K5V star indicated}
\end{figure*}

Despite being only 30 degrees from the Sun, the optical counterpart of A06200-00, subsequently designated V616 Mon, was rapidly located through its
nova-like behaviour  (Boley and Wolfson 1975), the accurate stellar position then allowing identification on the Palomar Observatory Sky Survey
charts (Ward \et\ 1975) with a m$_{B}$$\sim$20 star. Ward \et\ suggested the faint counterpart was a low mass, solar type star. A radio
counterpart was identified with the Mk 2 telescope at Jodrell Bank (Davis \et\ 1975), albeit delayed for a week as the telescope was in use by a guest observer.

A more detailed optical study, when the nova light had faded, confirmed the companion as a K5V star in a 7.8 hr period binary (McClintock \et\ 1983).
The small separation of the binary explained why Roche lobe overflow created an accretion disc around the compact companion, with a thermal-viscous disc
instability causing the X-ray and optical outburst.

Spectroscopic observation of the binary companion in quiescence revealed the presence of narrow absorption lines, showing an extremely large amplitude radial velocity 
(figure 6) which allowed McClintock and Remillard (1986) to 
derive a precise value for the binary period, and a Doppler-corrected spectrum of the  companion. The outcome was a 3$\sigma$ lower limit of 3.2 $\msun$
for the mass of the compact X-ray source, {\it independent} of distance and mass of the companion star. With reasonable assumptions regarding the K dwarf star, the
lower limit increased to $\sim$7.3 $\msun$, well above 
the maximum mass of a neutron star.

The low mass of  the optical counterpart to A0620-00 was critical in enabling  the mass of the
X-ray star to be determined with greater certainty than for Cygnus X-1, where the large and (at the time) uncertain mass of 
the companion supergiant led to residual doubts regarding the nature of the X-ray source (eg Trimble \et\ 1973).

\begin{figure}
\begin{center}
\hbox{
 \hspace{0.25 cm}
  \includegraphics[width = 11 cm ]{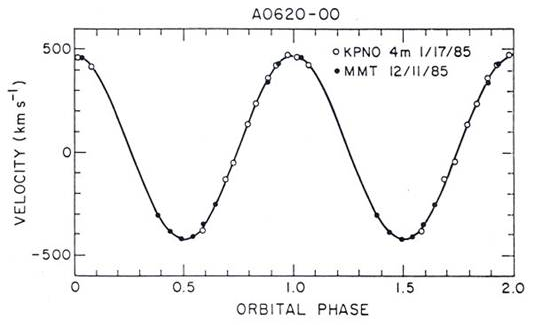}}
\end{center}
\caption{Radial velocity curve of the companion star to A0620-00. The extreme amplitude provided the strongest evidence to date of a black
hole (from Charles and Seward 1995)}       
\end{figure}

The return of A0620-00 has been forecast for circa 2033, based on the discovery from Harvard plates that V616 Mon previously flared up in 1917, although that
return date is subject to limitations in modelling the thermal-viscous limit cycle in the accretion disc. 
Meanwhile,
optical studies in quiescence remain of considerable interest, with the detection of rapid optical flaring, with rise times of 30 seconds or less, and 
a power-density
spectrum that may be characteristic of black hole binaries in their low state (Hynes \et\ 2003). Hynes \et\ concluded that
the flares are associated with the accretion flow rather than with an active companion, though it remains unclear whether they originate in the outer
disc, or are driven by events  in the inner region.

In another important step, using a SAS-3 spectrum taken in 1975 to estimate the radius of the innermost orbit, Gou \et\ (2010) have shown the black hole in 
A0620-00 to be spinning
quite slowly, with a spin parameter $a_*$ = 0.12$\pm$0.19, suggesting the radio jet seen in both flaring and quiescent states is probably disc driven.

Finally, a recent determination of the inclination of A0620-00 by Cantrell (2010), has allowed the black hole mass to be further refined to 6.6$\pm$0.25 $\msun$.

\section{Supermassive black holes in AGN}

A second important contribution from the Ariel 5 Sky Survey to the search for black holes was in establishing powerful X-ray emission  as a characteristic property  of
Seyfert galaxies, alongside the bright optical nucleus and broad permitted lines.   

Prior to the launch of Ariel 5 the majority of extragalactic X-ray source identifications were with rich galaxy clusters. Only NGC 4151 and 3C 273 had  been
identified uniquely with AGN in the 3U catalogue (Kellogg 1974). 60 Uhuru sources at latitude greater than 20 degrees remained unknown and it was suspected that the
majority of these unidentified high Galactic latitude sources (UHGLSs) might represent a new form of `X-ray Galaxy' (Giacconi 1973).  The difficulty of
identifying many unidentified sources at high Galactic latitude, in both Uhuru and Ariel 5 catalogues, was again due to positions known only to a few tenths
of a sq. deg.

\begin{figure}
\begin{center}
\hbox{
\hspace{0.5 cm}
  \includegraphics[width=11 cm]{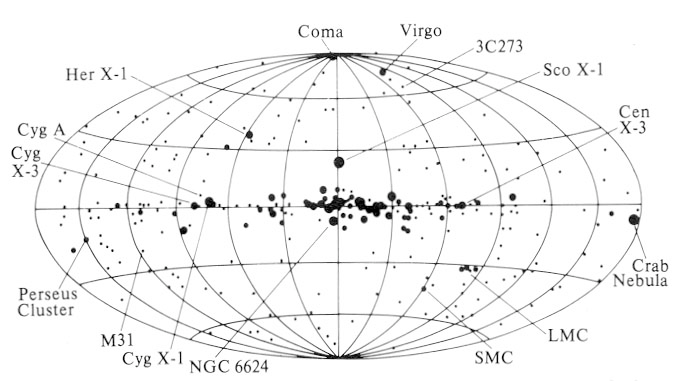}}
\end{center}
\caption{Map of 297 X-ray sources detected with the Ariel 5 3A catalogue, plotted in Galactic coordinates with source diameter proportional to the log of
the X-ray flux. Many of the faint sources at high Galactic latitude were identified with Seyfert
galaxies and galaxy clusters}
\end{figure}

The Seyfert galaxy NGC3783 was the first new Ariel 5 identification (Cooke \et\ 1976), being the brightest object in the 2A1135-373 error box. Optical
spectra obtained with the 3.8m Anglo Australian Telescope (AAT) revealed the presence of [FeX] and other high ionisation lines, making the X-ray association more likely. Nine further
coincidences of bright Seyferts with Ariel 5  sources quickly followed, enabling a report at the 1976 Relativistic Astrophysics meeting in
Boston (Pounds
1977) that those 10 Seyfert identifications, together with 21 new rich galaxy cluster/X-ray identifications, had essentially solved the mystery of the `UHGLSs'.
How a combination of modest gains in source location and in source detection sensitivity allowed the SSI to achieve that breakthrough is recalled by a
key participant at the time (Elvis 2012). 

A subsequent search of Ariel 5 X-ray error boxes led to the discovery of several previously unknown Seyfert galaxies (Ward \et\ 1978), while other
Seyfert identifications were found in more precise  SAS-3 error boxes (Schnopper \et\ 1977, 1978) and from analysis of the final 4U source catalogue 
(Tananbaum \et\ 1978).

Establishing powerful X-ray emission as a characteristic property of Seyfert galaxies was further strengthened when Elvis \et\ (1978) showed, for  a
sample of 15 Seyfert galaxies, that the X-ray luminosity was correlated with the infrared and optical luminosity and with the width of the broad emission
lines, but not with the radio flux, strongly suggesting a common origin in the innermost  $\leq$0.1 pc.

Although it was widely believed that the X-ray emission was produced by accretion onto a supermassive object (see Rees, Begelman and Blandford 1981 for a review), 
confirmation that it was the signature of a supermassive black hole required more information, in particular on the compactness of the emission region.
An initial search for sufficiently rapid X-ray variability was disappointing, with a sample of 38 bright AGN observed by HEAO-2 finding relevant variability on a timescale
less than 12 hours (Tennant and Mushotzky 1983).

\begin{figure*}
\begin{center}
\hbox{
 \hspace{0.5 cm}
  \includegraphics[width=5 cm]{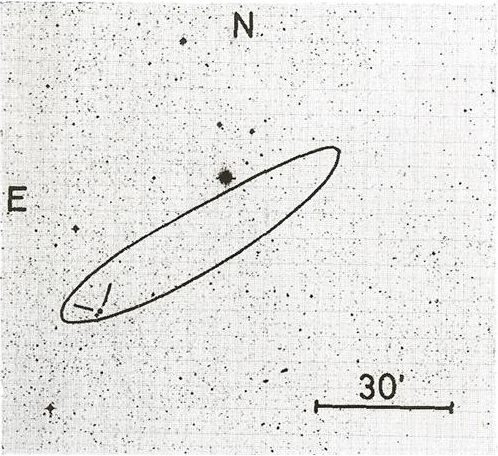}
 \hspace{0.5 cm}
  \includegraphics[width = 5 cm, angle=360]{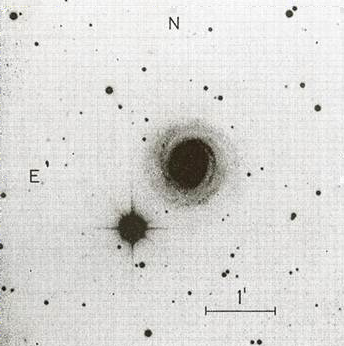}}
  \end{center}
\caption{(left) 90 $\%$ confidence error box of 2A 1135-373 superimposed on a UK Schmidt Telescope plate. The Seyfert galaxy NGC 3783 is indicated.
(right) Blue filter 3.8m AAT plate showing the Seyfert galaxy NGC 3783, the first of a new class of X-ray emitters identified with the Ariel 5 SSI}
\end{figure*}

\begin{figure}
\begin{center}
\hbox{
 \hspace{1.5 cm}
  \includegraphics[width = 9 cm ]{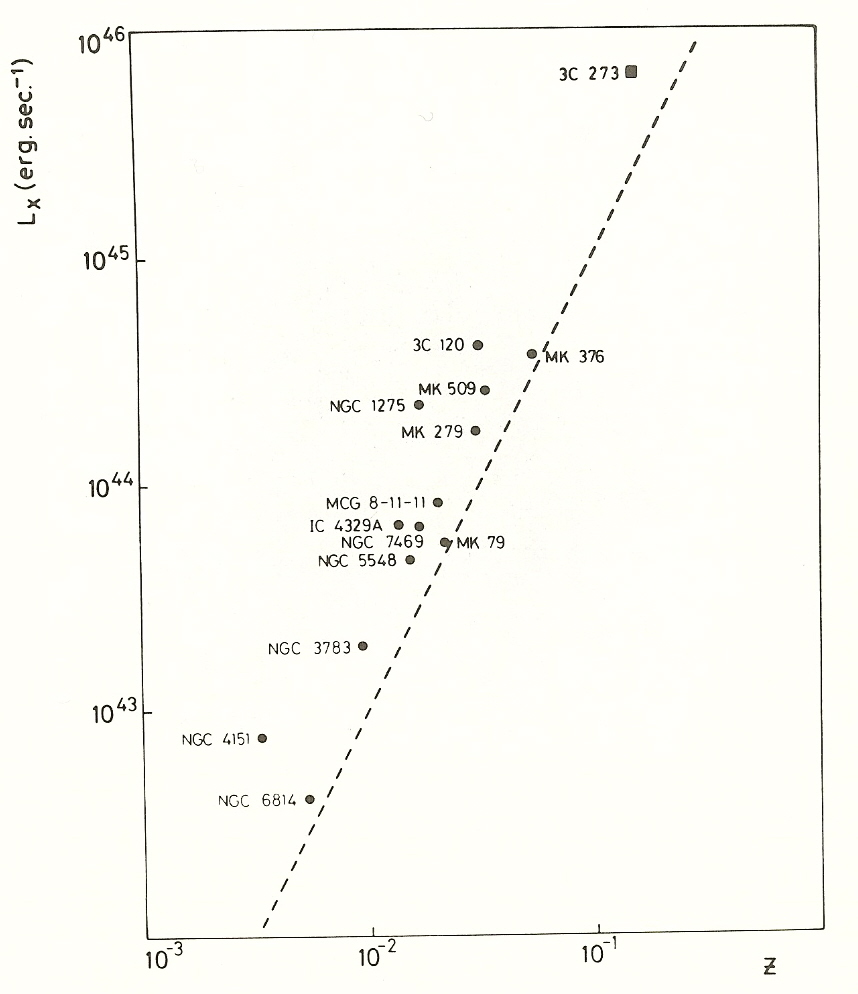}}
\end{center}
\caption{Luminosity-redshift plot of 13 Seyfert galaxies (10 new) identified in the 2A catalogue, together with the quasar 3C 273}
\end{figure}

That crucial next step came, however, from uniquely long uninterrupted observations afforded by the highly eccentric orbit of the ESA satellite
EXOSAT (Pallavicini and White 1988). Operational from 1983-86, it was only in the final months when ESA was persuaded to approve observations lasting for a
full $\sim$4 day spacecraft orbit. Several bright AGN were prime targets, since Ariel 5 had found evidence for variability on timescales of days (Marshall
\et\ 1981).
The outcome was remarkable, finding the X-ray emission from several Seyfert galaxies to vary, with large amplitude (ie coherently), on hours or less 
(Pounds and McHardy 1988). Light travel time
arguments showed the X-ray sources to be very compact, with the clear implication that Seyfert X-ray sources were more massive analogues of accreting stellar mass 
black holes.

\begin{figure}
  \includegraphics[width = 12 cm ]{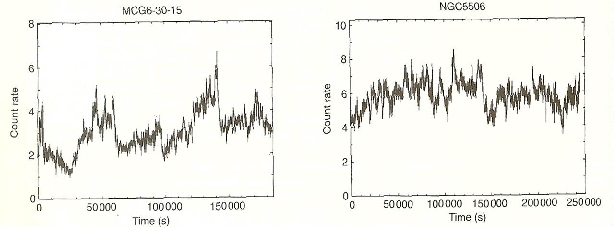}
\caption{X-ray light curves of two Seyfert galaxies with rapid, high amplitude variability providing a compelling argument for a supermassive black hole}
\label{fig:5}       
\end{figure}

Further evidence supporting that conclusion came from the discovery of `X-ray reflection' from observations
with the Japanese GINGA satellite (1987-90), in which spectral features were found which appeared to arise from scattering of X-rays off optically thick matter,
probably the accretion
disc (Pounds \et\ 1990, Nandra and Pounds 1994). The following Japanese satellite, ASCA, obtained higher resolution spectra which showed the accompanying
fluorescent Fe K line to have a broad red wing (Tanaka \et\ 1995, Nandra \et\ 1997), interpreted as a gravitational red shift and inspiring a major and continuing research effort to explore the effects of
strong gravity in the inner parts of the accretion disc (Fabian \et\ 2000).

\section{ULXs and intermediate mass black holes}

The discovery of several ultra-luminous X-ray sources in nearby galaxies (Fabbiano 1989) has suggested the possible existence of a third class of black hole, with a mass intermediate
between the now well-established categories of stellar and supermassive black holes. With X-ray luminosities, assuming isotropic emission, up to
$\sim$$10^{41}$ erg s$^{-1}$, the implied black hole mass accreting at the Eddington limit would be $\sim$$10^{3}$ $\msun$. The existence of such
intermediate mass black holes remains unclear, however, with a strong possibility being that the X-ray flux is enhanced by being beamed in our direction (King \et\ 2001).

SS433 is an example of a stellar mass black hole with X-ray emission known to be beamed away from the line of sight. In the context of this
review it is interesting to note that SS433 remained largely ignored until flagged up by the adjacent Ariel 5 source A1909+04  (Seward \et\ 1976). Follow-up observations 
at the AAT by Clark and Murdin (1978) found a strong emission line spectrum, with several lines they could not identify. Margon 
(1979) made further observations at the Lick Observatory, finding the unidentified lines to shift in wavelength from night to night. Over the following months it emerged
that SS433 is a truly remarkable object, with a pair of jets travelling at relativistic speeds and precessing about a common axis with a 164 day period (Fabian and Rees 1979). While 
SS433 stands out as an example of extreme super-Eddington accretion in a black hole binary, it can also be seen as a ULX
viewed side-on.

\section{Summary and update} 

Establishing that black holes exist with masses in the range $\sim$5-15 $\msun$ and $\sim$$10^{6} - 10^{9}$ $\msun$ has been one of the major
scientific returns from space observations. The high luminosity and powerful outflows from black holes, when amply fed, offers exciting prospects for
studying the physics of accretion processes that play such a wide role in astronomy, while exploring matter in regions of strong gravity.

Looking back over 50 years of X-ray astronomy the discovery of A0620-00 still stands out, only the Magnetar SGR 1806 (Palmer \et\ 2005) having exceeded its peak X-ray flux.
Given the limited sensitivity and infrequent deployment of all-sky X-ray detectors, similar dramatic events may have been missed. The GINGA transient GS2000+25
(Tsunemi \et\ 1989) appears to have been similar to A0620-00, but more distant and a factor 6 less bright. Just weeks before this Workshop, on 16 September 2012,
the SWIFT GRB monitor triggered on a source in Scorpius which 
became as
bright as the Crab nebula 2 days later, with early indications of being a further `once in a mission event' (Neil Gehrels in NASA PR). However, the conclusion must
be that such events are indeed quite rare. 

Ozel \et\ (2010) review the mass distribution of Galactic black hole candidates, including 23 confirmed black hole binaries. 
Dynamical data show 17 of those to have low mass optical companions
with a narrow distribution of black hole masses, of 7.8$\pm$1.2 $\msun$, for the best determined
systems. A significant absence of black hole masses between 5$\msun$ and the neutron star mass limit of 2$\msun$ was a surprise, which Ozel \et\ speculate may be due
to a sudden fall in the energy of the supernova explosion with increasing 
progenitor mass.

Binary periods for the low mass transient systems range from $\sim$4 hr to $\sim$6 d (an exception being GRS 1915+105), and $\sim$32 hr to $\sim$5.6 d for the persistent 
high mass sources, comfortably meeting the requirement anticipated in the Introduction for copius mass transfer by Roche lobe overflow. 
Such powerful X-ray sources will be the exception, of course, with the vast majority of Galactic black holes being limited to inefficient Bondi
accretion from the local
ISM, and much too faint to be detected.

Since launch in 2004 the NASA Swift satellite has been detecting $\sim$100 Gamma Ray Bursts (GRB) each year (Gehrels and Meszaros 2012), with the few-arc-sec
positions obtained by detecting the X-ray afterglow allowing many bursts to be optically identified. Many GRB are found to be associated with
supernovae in galaxies at high redshift. With a likely origin in the collapse of a massive star (MacFadyen and Woosley 1999), the remarkable
inference is that Swift is observing the birth of stellar mass black holes from across the Universe, at a rate - allowing for beaming and other
factors - of $\sim$$10^{5}$ a year.

One unusually long event observed by Swift in March 2011 has been interpreted as the capture of a star by a SMBH (Burrows \et\ 2012, Levan \et\ 2012), 
prefacing an important new way of studying timescales in the accretion disc of an AGN. Follow-up X-ray observations with \suzaku\ and \xmm\ detected Quasi Period
Oscillations (QPO), with a transient periodicity at 3.5 minutes, evidently relating to processes very close to the innermost stable circular orbit (Reis \et\ 2012).
That transient event adds to the first reliable QPO detection in an AGN (Gierlinski \et\ 2008), providing a further example of the similarities in disc physics shared 
by accreting black holes in X-ray binaries and AGN. 

The interpretation of
flickering in the emission from Sgr A as the disruption of passing asteroids (Zubovas \et\ 2012) provides a further indication that the feeding of SMBH
will range from short sub-Eddington events to less frequent minor and major mergers.
However, mergers seem more likely to have led to the heavily dust-obscured galaxies (DOGS) found in the WISE sky survey (NASA PR 29/8/12); NuStar observations should
confirm whether these are super-Eddington AGN.

\begin{figure*}
\begin{center}
\hbox{
 \hspace{0.5 cm}
\includegraphics[width=5.5cm]{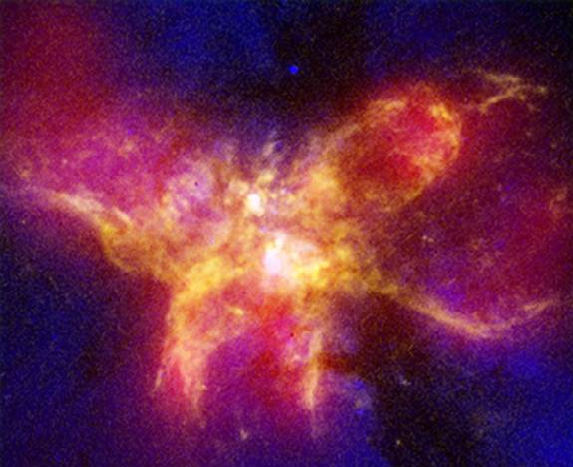}
 \hspace{0.5 cm}
  \includegraphics[width=4.5cm, angle=360]{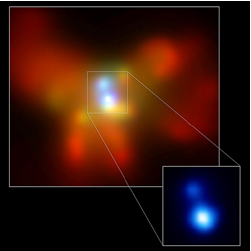}}
 \end{center}
\caption{Optical and \chandra\ images of NGC 6420, believed to be a pair of merging galaxies. The two bright point-like X-ray
sources in the right hand panel are probably SMBH at the respective nuclei}
\end{figure*}

The study of supermassive black holes has been a major beneficiary of three powerful X-ray observatories, \chandra, \xmm\ and \suzaku, launched since  1999. Long
exposures on HST deep fields have shown, via coincident X-ray emission, that powerful X-ray emission - and implicitly the presence of supermassive black holes - extends
to AGN at high redshift. The high angular resolution of \chandra\ resolved two point-like X-ray sources near the centre of NGC 6420 (figure 11), believed to be a pair of
merging galaxies (Komossa \et\ 2003). The eventual merger of the SMBHs will be a spectactular event,  with a substantial fraction of the combined rest mass released over
a few hours, mainly as a QPO gravity wave outburst.  

Finally, as discussed elsewhere at this Workshop, highly ionised outflows with velocities $\sim$0.1-0.2c were detected in X-ray spectra of non-BAL AGN a decade ago
(Pounds \et\ 2003, Reeves \et\ 2003). Subsequent observations (Cappi 2006) and searches of the \xmm\ (Tombesi \et\ 2010, 2011) and \suzaku\ (Gofford \et\ 2013) data
archives have shown such ultra-fast outflows to be a common feature of luminous AGN, indicating that powerful black hole winds are likely to have a wider importance in galaxy feedback
models.

\begin{acknowledgements} The outstanding efforts of many colleagues and former students enabled the X-ray Astronomy Group at Leicester to play a significant role in 
establishing the existence of black holes in both stellar systems and in the nuclei of Active Galaxies. That opportunity was, of course, only possible by building on the 
pioneering achievements of other scientists and engineers who have established X-ray astronomy as a vibrant discipline in Space Science.
Grateful thanks are also due to Jeff McClintock, Paul Gorenstein and Martin Ward who provided valuable comments on the draft manuscript. 
\end{acknowledgements}

\end{document}